\documentclass[pdflatex,sn-mathphys-num]{sn-jnl}

\usepackage{graphicx}%
\usepackage{multirow}%
\usepackage{amsmath,amssymb,amsfonts}%
\usepackage{amsthm}%
\usepackage{mathrsfs}%
\usepackage[title]{appendix}%
\usepackage{xcolor}%
\usepackage{textcomp}%
\usepackage{manyfoot}%
\usepackage{booktabs}%
\usepackage{algorithm}%
\usepackage{algorithmicx}%
\usepackage{algpseudocode}%
\usepackage{listings}%

\theoremstyle{thmstyleone}%

\theoremstyle{thmstyletwo}%

\theoremstyle{thmstylethree}%

\begin{document}

\title[Routing protocols for AI multi-agent systems]{Adaptive routing protocols for determining optimal paths in AI multi-agent systems: a priority- and learning-enhanced approach}

\author[1]{\fnm{Theodor} \sur{Panayotov}}\email{t@ethermind.ai}
\author*[2,3]{\fnm{Ivo} \sur{Emanuilov}}\email{ivo.emanuilov@kuleuven.be}
\equalcont{The authors contributed equally to this work.}

\affil*[1]{\orgname{Ethermind Ltd}, \orgaddress{\street{71-75 Shelton Street, Covent Garden}, \city{London}, \postcode{WC2H 9JQ}, \country{United Kingdom}}}

\affil[2]{\orgdiv{imec - Centre for IT \& IP Law}, \orgname{KU Leuven}, \orgaddress{\street{6 Sint-Michielsstraat, bus 3443}, \city{Leuven}, \postcode{3000}, \country{Belgium}}}

\affil[3]{\orgname{GATE Institute - University of Sofia}, \orgaddress{\street{5 James Bourchier Blvd}, \city{Sofia}, \postcode{1164}, \country{Bulgaria}}}


\abstract{As distributed artificial intelligence (AI) and multi-agent architectures grow increasingly complex, the need for adaptive, context-aware routing becomes paramount. This paper introduces an enhanced, adaptive routing algorithm tailored for AI multi-agent networks, integrating priority-based cost functions and dynamic learning mechanisms. Building on an extended Dijkstra-based framework, we incorporate multi-faceted parameters such as task complexity, user request priority, agent capabilities, bandwidth, latency, load, model sophistication, and reliability. We further propose dynamically adaptive weighting factors, tuned via reinforcement learning (RL), to continuously evolve routing policies based on observed network performance. Additionally, heuristic filtering and hierarchical routing structures improve scalability and responsiveness. Our approach yields context-sensitive, load-aware, and priority-focused routing decisions that not only reduce latency for critical tasks but also optimize overall resource utilization, ultimately enhancing the robustness, flexibility, and efficiency of multi-agent systems.}


\maketitle

\section{Introduction}\label{sec1}

As AI multi-agent systems (MAS) scale, agents must rapidly coordinate and allocate tasks among heterogeneous nodes. Modern agent-based networks often involve diverse agents managing advanced large language models (LLMs), processing high volumes of data, and serving requests of varying importance. Ensuring that critical, time-sensitive tasks are routed through efficient, reliable paths is essential for maintaining system performance and satisfying stringent user demands.

Traditional routing algorithms, while robust for general-purpose data networks, rarely leverage the semantic richness of AI environments. They often ignore factors such as the sophistication of the agent model, the current load, or user-defined priorities. Furthermore, static weighting approaches fall short in dynamic conditions, where sudden workload shifts, agent failures, or bandwidth constraints can drastically alter optimal routing paths.

In this paper, we introduce an adaptive priority-based Dijkstra's algorithm (APBDA) for multi-agent AI systems. Our contributions go in four main directions:
    \begin{itemize}
        \item \textbf{Priority-based and context-aware costs}. We incorporate a multifactor cost function considering task complexity, user priority, agent capability, availability, bandwidth, latency, model sophistication, and reliability.
        \item \textbf{Adaptive weighting via reinforcement learning (RL)}. Instead of relying on fixed weights, we propose using RL techniques to dynamically adjust the weighting factors, $w_i$. By continuously monitoring system performance, the algorithm learns optimal weighting configurations over time.
        \item \textbf{Heuristic filtering}. To reduce computational complexity in large-scale networks, we introduce heuristic filters that prune suboptimal candidate paths early, speeding up route discovery.
        \item \textbf{Hierarchical routing structure}. We suggest an optional hierarchical overlay, where clusters of agents are grouped, and routing occurs both within and between these clusters. This approach facilitates scalable deployments in large MAS settings
    \end{itemize}

The proposed solution adapts quickly to evolving conditions, ensuring that critical requests receive preferential treatment without sacrificing overall resource efficiency. This framework positions the algorithm as a powerful tool for next-generation distributed AI systems.

\section{Related work}\label{sec:related_work}

Foundational shortest-path algorithms like Dijkstra’s \cite{Dijkstra1959} and A* \cite{Hart1968} have long been used in network routing and multi-agent path planning. Prior work on multi-agent task allocation considers high-level heuristics for distribution and scheduling \cite{Lesser1995,Stone2000}, while distributed machine learning frameworks (e.g., parameter servers, federated learning) emphasize bandwidth and latency optimization \cite{Li2020}. Others have focused on consensus algorithms which ensure that agents can communicate and agree on a common state or value, which is essential for effective path planning and routing in dynamic environments \cite{amirkhani2022consensus}. With the large-scale deployment of LLMs, so-called LLM-based multi-agent systems also gained popularity. Research on agent communication in LLM-based MAS has considered various communication paradigms (eg, cooperative, debate, and competitive), structure (eg, the use of layered communication structures and shared message pools to enhance coordination and task allocation) and content (eg, in the form of text)\cite{guo2024survey}. Recently, adaptive routing solutions have emerged, incorporating machine learning-based approaches for dynamic traffic management \cite{Kleinberg2007}.

However, these methods often lack holistic integration of AI-centric parameters. Few, if any, incorporate model sophistication, agent reliability histories, or RL-based adaptation of the routing parameters. By integrating these advanced considerations, our approach expands the scope and flexibility of adaptive routing in AI MAS.

\section{Algorithm for calculating costs between agents}\label{sec2}

To reflect real-time conditions and system goals, we define cost variables as follows:

\begin{itemize}

    \item \textbf{Task complexity (T)}, ie, computational resources required for the given task.
    \item \textbf{User request priority (P)}, ie, higher values for more urgent or critical requests.
    \item \textbf{Agent processing capability ($C_i$)}, ie, performance metrics of agent $i$ (e.g., FLOPS).
    \item \textbf{Machine availability ($A_i$)}, ie, readiness to accept new tasks, reflecting workload and uptime.
    \item \textbf{Bandwidth ($B_{ij}$)}, ie, available network bandwidth between agents $i$ and $j$.
    \item \textbf{Latency ($L_{ij}$)}, ie, network delay between agents $i$ and $j$.
    \item \textbf{Model sophistication ($M_i$)}, ie, quality and complexity of models running on agent $i$.
    \item \textbf{Load factor ($F_i$)}, ie, current utilization and system load of agent $i$.
    \item \textbf{Reliability ($R_i$)}, ie, historical stability and failure rates of agent $i$.
\end{itemize}


We define the cost of moving from agent $i$ to agent $j$ as the following adaptive cost function:

\[
\text{Cost}_{ij} = w_1 \left(\frac{T}{C_j}\right) + w_2 \left(\frac{P}{A_j}\right) + w_3 \left(\frac{P}{B_{ij}}\right) + w_4 (P \cdot L_{ij}) + w_5 \left(\frac{F_j}{C_j}\right) + w_6 \left(\frac{1}{M_j}\right) + w_7 \left(\frac{1}{R_j}\right)
\]

Each $w_i$ initially reflects system-level policy priorities. For example, $w_4$ may be larger in latency-sensitive scenarios. Over time, these weights are adjusted via RL-based methods (eg, Q-learning or policy gradients), allowing the system to learn optimal policies from operational feedback (eg, task completion times, agent utilization metrics).

\textbf{Task complexity and processing capability} ($T/C_j$) refers to the idea that higher complexity tasks are better handled by agents with larger $C_j$. Low $C_j$ with high $T$ leads to higher costs.
\textbf{User priority and availability} ($P/A_j$) reflects the idea that high-priority requests should target agents that are readily available.
\textbf{Priority and bandwidth} ($P/B_{ij}$) refers to the the need for critical data should flow through higher-bandwidth links to minimise transfer delays.
\textbf{Priority and latency} ($P \cdot L_{ij}$) reflects the idea that latency-sensitive requests heavily penalise routes with large $L_{ij}$.
\textbf{Load factor and processing capability} ($F_j/C_j$) represents the idea that agents already under heavy load are less attractive, especially if their capability does not compensate.
Model sophistication ($1/M_j$) is concerned with the idea that more sophisticated models are generally preferred, particularly if the task complexity demands advanced inference.
Finally, \textbf{reliability} ($1/R_j$) concerns the idea that routes passing through less reliable agents incur greater risk and thus higher costs.

\section{Enhanced adaptive priority-based Dijkstra's algorithm (APBDA)}

We extend Dijkstra’s algorithm as follows:

\begin{enumerate}
    \item Initialise $\text{TotalCost}[v] = \infty$ for all agents $v$, and $\text{TotalCost}[s] = 0$ for source $s$.
    \item Maintain a min-priority queue keyed by $\text{TotalCost}$.
    \item Iteratively extract the minimum-cost agent $u$. For each neighbor $v$:
    \begin{itemize}
        \item Compute $\text{Cost}_{uv}$ based on the current weights $w_i$, which are periodically updated through RL.
        \item If $\text{TotalCost}[u] + \text{Cost}_{uv} < \text{TotalCost}[v]$, update $\text{TotalCost}[v]$ and set $\text{Predecessor}[v]=u$.
    \end{itemize}
    \item Stop once the destination $d$ is reached or the queue is empty.
\end{enumerate}

To continuously improve routing decisions, we employ a reinforcement learning (RL) mechanism. In this mechanism, \textbf{state representation} includes network-wide statistics such as average latency, agent load distribution, recent reliability incidents, and priority profiles of incoming tasks. \textbf{Action space} corresponds to adjusting the vector $w = (w_1, w_2, \ldots, w_7)$. Small perturbations in weights can shift routing decisions. The \textbf{reward function} is a combination of system-level metrics, eg, inverse of average completion time for high-priority tasks, balanced with load distribution fairness and agent reliability. High rewards occur when the routing decisions lead to timely completion of critical requests and well-distributed workloads.

Over time, the RL algorithm converges to an effective weighting policy that improves both global performance and responsiveness to changing conditions.

\section{Heuristic filtering and hierarchical routing}

In large-scale MAS, considering every agent or link may be impractical. We propose two enhancements:

The first is called \textbf{heuristic filtering}. Before running APBDA, we can prune obviously suboptimal paths using heuristics—eg, eliminate edges with consistently high latency or agents with persistently low reliability from the initial search space. These filters reduce computational overhead.

The second is called \textbf{hierarchical clustering}. Group agents into clusters, each managed by a cluster-head. Inter-cluster routing uses aggregated metrics to select promising routes, and intra-cluster routing applies the full APBDA. This multi-level approach improves scalability and reduces the complexity of global routing decisions.

\section{Pseudocode implementation}

    
    
    


The following lines represent a pseudocode implementation of the proposed algorithm.
\begin{algorithmic}[1]
\Function{AdaptivePriorityDijkstra}{graph, source, destination, task\_complexity, priority\_level, weights}
    \ForAll{vertex $v$ in graph}
        \State $TotalCost[v] \gets \infty$
        \State $Predecessor[v] \gets \text{None}$
    \EndFor
    \State $TotalCost[source] \gets 0$
    \State priority\_queue $\gets$ min-priority queue ordered by $TotalCost$
    \State priority\_queue.insert(source)

    \Comment{Heuristic filtering step (optional)}
    \State \textbf{/* filtered\_graph $\gets$ apply\_heuristics(graph, weights) */}

    \While{priority\_queue is not empty}
        \State $u \gets$ priority\_queue.extract\_min()
        \If{$u$ == destination}
            \State \textbf{break}
        \EndIf
        \ForAll{neighbor $v$ of $u$}
            \State $Cost\_uv \gets$ \Call{compute\_cost}{$u, v, task\_complexity, priority\_level, weights$}
            \State $Alt \gets TotalCost[u] + Cost\_uv$
            \If{$Alt < TotalCost[v]$}
                \State $TotalCost[v] \gets Alt$
                \State $Predecessor[v] \gets u$
                \If{$v$ in priority\_queue}
                    \State priority\_queue.decrease\_key($v, Alt$)
                \Else
                    \State priority\_queue.insert($v$)
                \EndIf
            \EndIf
        \EndFor
    \EndWhile
    \State \Return \Call{reconstruct\_path}{Predecessor, source, destination}
\EndFunction

\Function{compute\_cost}{$u, v, T, P, w$}
    \State $C_v \gets$ processing\_capability($v$)
    \State $A_v \gets$ availability($v$)
    \State $B\_uv \gets$ bandwidth($u, v$)
    \State $L\_uv \gets$ latency($u, v$)
    \State $F_v \gets$ load\_factor($v$)
    \State $M_v \gets$ model\_sophistication($v$)
    \State $R_v \gets$ reliability($v$)
    \State $(w1, w2, w3, w4, w5, w6, w7) \gets w$
    \State $Cost\_uv \gets w1 \cdot \frac{T}{C_v} + w2 \cdot \frac{P}{A_v} + w3 \cdot \frac{P}{B\_uv}$
    \State \hspace{0.45cm} $+ w4 \cdot (P \cdot L\_uv) + w5 \cdot \frac{F_v}{C_v}$
    \State \hspace{0.45cm} $+ w6 \cdot \frac{1}{M_v} + w7 \cdot \frac{1}{R_v}$
    \State \Return $Cost\_uv$
\EndFunction
\end{algorithmic}
\section{Demonstration of adaptive routing}

Under sudden\textbf{ high-priority demands}, the APBDA focuses on minimizing latency and ensuring sufficient bandwidth. After several iterations, RL may increase $w_4$ (latency factor) and $w_3$ (bandwidth factor) to direct traffic toward agents with reliable, fast links.

For \textbf{lower-priority tasks}, RL might gradually reduce the latency emphasis and rely more on maximizing agent availability or reliability. This leads to more balanced resource utilization and allows prime resources to be conserved for urgent tasks.

There are several advantages to our enhanced approach. It enables \textbf{contextual intelligence}. By integrating multiple factors—capability, load, reliability—the algorithm selects more intelligent routes tailored to the nature of each request. It also affords \textbf{dynamic adaptation}. RL-based weight adjustments ensure that the routing strategy evolves with changing network conditions, maintaining optimal performance over time. Heuristic filtering and hierarchical clustering significantly reduce \textbf{computational complexity} for large-scale deployments, thereby enabling better scalability. Finally, incorporating \textbf{reliability} and model sophistication reduces the risk of route failures and suboptimal allocations.

\section{Discussion and future work}

Although the APBDA represents a significant step toward adaptive, intelligent routing, several areas remain open for exploration. Integrating trust mechanisms or security requirements into the cost function could ensure that sensitive data only travels through trusted agents. Energy efficiency considerations, particularly for battery-operated agents (ef, in swarm robotics), could be added as another factor. 

Future research can explore more sophisticated RL methods (eg, distributed multi-agent RL) to handle weighting factor updates in a decentralized manner. Further, comprehensive simulations and real-world deployments in federated learning platforms, distributed robotics, or large-scale IoT networks would validate the approach’s effectiveness and guide refinements.

\section{Conclusion}

This paper introduced an adaptive priority-based Dijkstra’s Algorithm (APBDA) for routing in AI multi-agent systems. By integrating a rich, priority-aware cost function, employing RL-driven adjustments to weighting factors, and supporting heuristic and hierarchical strategies, the algorithm dynamically adapts to complex and evolving network conditions. The result is an intelligent, robust, and scalable solution that ensures timely responses to high-priority tasks while efficiently managing global resources.

\bmhead{Acknowledgements}

We thank colleagues and reviewers for their constructive feedback and suggestions that shaped the development of these adaptive routing strategies.

\end{document}